\documentclass[vecphys]{svmult}

\usepackage{makeidx}         % allows index generation
\usepackage{graphicx}        % standard LaTeX graphics tool
\usepackage{multicol}        % used for the two-column index
\usepackage[bottom]{footmisc}% places footnotes at page bottom
\makeindex             % used for the subject index
\def\kmskpc{\ km s$^{-1}$ kpc$^{-1}$}

\begin{document}

\title*{Local Phase Space - Shaped by Chaos?}
\author{Dalia Chakrabarty\inst{1}}
\institute{School of Physics $\&$ Astronomy, University of Nottingham, Nottingham NG7 2RD, U.K.
\texttt{dalia.chakrabarty@nottingham.ac.uk}}
\maketitle

\section{Introduction}
\label{sec:1}
% Always give a unique label
% and use \ref{<label>} for cross-references
% and \cite{<label>} for bibliographic references
% use \sectionmark{}
% to alter or adjust the section heading in the running head
%Your text goes here. Use the \LaTeX\ automatism for your citations
%\cite{monograph}.

The exploration of the nature of the phase space that we live in, is a
naturally attractive endeavour. On astronomical scales, this
translates to an exercise in understanding the state of the phase
space in the neighbourhood of the Sun. Of course, for a long time,
this was impeded by the dearth of sufficiently large samples of proper
motion data of the nearby stars. However, with the information
available in Woolley's catalogue \cite{woolley}, Agris Kalnajs
concluded the local velocity space to be basically bimodal
\cite{agris} and attributed this behaviour to the rough proximity of
the Sun to the Outer Lindblad Resonance of the central bar in the
Galaxy ($OLR_b$). The observational domain expanded with the
availability of the transverse velocity information of stars in the
vicinity of the Sun, as measured by {\it Hipparcos}. Since then many
workers have attempted to chart the local phase space distribution
$DF$ \cite{jameswalter, fux_aa, skuljan} and investigate the origin of
the structure in the same \cite{fux_aa, walterolr, tremainewu,
quillen, famaey05}.

Here we pursue the hypothesis that the observed state of the
local phase space owes to the dynamical effect of Galactic features
such as the central bar or the spiral pattern. However, dynamical
modelling of the Solar neighbourhood is incomplete without the
inclusion of the effects of {\it both} the bar and the spiral pattern;
after all, a major resonance of the bar occurs in the vicinity of the
Sun, (as suggested by the Kalnajs Mechanism) and the inter-arm
separation in the Galactic spiral pattern falls short of the average
epicyclic excursion of a star at the Solar radius
\cite{chakrabarty_aa}. 

However, this joint perturbation has been considered only
occasionally \cite{chakrabarty_aa, quillen}. That too, it is
contentious if the Galactic spiral arms should attach themselves to
the ends of the bar or rotate with a pattern speed ($\Omega_s$) that
is distinct from that of the bar ($\Omega_b$).  The former scenario
was invoked to produce self-consistent models of the galaxies NGC
3992, NGC 1073, and NGC 1398 \cite{kaufmann96} while the best-fit
model for NGC 3359 was reported to manifest distinct pattern speeds
for the bar and the spiral \cite{boon05}. Such dissimilar pattern
speeds were considered for the first time by \cite{chakrabarty_aa} in
her modelling of the Solar neighbourhood. The effect of the relative
separation between the locations of the resonances due to the Galactic
bar and the spiral pattern is expected to bear interesting dynamical
consequences.

In this article, we report the results of modelling of the local phase
space, undertaken with the aim of disentangling the mystery about the
origin of the observed phase space structure. In particular, we
address the question of the imperativeness of the presence of chaos,
in order to explain this structure.

\section{Local Phase Space}
\noindent
The exploration of the physics responsible for the observed $DF$,
follows the exercise of density estimation in phase space, given the
observed velocity data. Assuming the Galactic disk to be ideally flat,
we approximate the phase space as given by four coordinates: the two
plane of the disk spatial coordinates (radius $R$ and azimuth $\phi$)
and the (heliocentric) radial and transverse velocities ($U$ and $V$,
respectively). $U$ is measured positive in the direction of the
Galactic centre while $V$ is positive along the sense of Galactic rotation.
Thus, our simulated velocity space is 2D in nature.

The stellar $U-V$ distribution is noted to be highly non-linear and
multi-modal. \cite{chakrabarty_aa} uses the same distribution that was
used in \cite{fux_aa}; this construction uses $V$ of single stars with
distance $d < 100$ pc and dispersion of parallax $\sigma(\pi) <
0.1\pi$ from the Hipparcos Catalogue and $U$ of the 3481 stars in the
{\it Hipparcos} Input Catalogue. This measured kinematic data, when
smoothed by the bisymmetric adaptive kernel estimator, (as used by
\cite{skuljan}) suggests five major clumps in the local velocity $U-V$
plane (see Figure~2 in \cite{chakrabarty_aa}); these have been identified
with the Hercules, Hyades, Pleiades, Coma Berenecius and Sirius
stellar streams or moving groups. It is the Hercules stream that
corresponds to the smaller mode while the bigger mode constitutes the
other four groups.

It is accepted that other choices of the kernel might have given rise
to a different picture of the $U-V$ plane. This is expected to be more
the case at the outer parts of the distribution, where there are
relatively fewer stars than near the central parts (between -50 km
s$^{-1} < V <$ 10 km s$^{-1}$ and -50 km s$^{-1} < U <$ 50 km
s$^{-1}$).  To check if the suggestion of the five streams is an
idiosyncracy of the density estimation procedure, other kernel
estimators were tried; it was concluded that this did not affect the
above mentioned central part of the distribution in question, though
the outer parts were indeed affected.

It merits mention that in the course of the density estimation
procedure, as suggested by \cite{skuljan}, the calculation of the
local bandwidth $\lambda_i$, at the $i^{th}$ velocity data point,
involves achieving convergence, by iteratively toggling between the
distribution estimate at this point and $\lambda_i$. The effect of
varying the initial seed for the density estimate, on the result of
this iteration, has however not been thoroughly investigated
or reported; subsequent workers \cite{fux_aa, chakrabarty_aa} have
used values provided by \cite{skuljan}.

Thus, accepting that the observed structure in the local velocity
space has been robustly estimated, test particle simulations (2-D by
nature) were carried out to constrain the model that best fit the
observations.

\section{Simulations}
\noindent
The simulations are test-particle in nature. This form of constricted
N-body sampling of the stars is sufficient in the outer parts of disk
galaxies and is useful when there is a multi-dimensional parameter
space that need to be scanned. In the simulations, a bunch of initial
phase space coordinates ($\sim$3500, i.e. the number of stars used to
construct the local $U-V$ diagram) are evolved in the potential of the
Galactic disk, as perturbed by the potentials of the bar and/or the
spiral pattern. The disk is assigned an initial phase space
distribution function given by the doubly cut-out forms used in
\cite{wynjenny}; this ensures an exponential surface mass density
profile. We choose to ascribe a Mestel potential to the disk, in order
to recover a flat rotation curve; the warmth parameter in this
potential is maintained sufficiently high to ensure the values of
radial and transverse velocity dispersions and vertex deviations of
stars in the vicinity of the Sun. The choice of the perturbation
potential is motivated by their geometries: the bar is modelled as a
rigidly rotating quadrupole while the spiral potential is approximated
by a logarithmic spiral.

\subsection{Spiral Characteristics}
\noindent
The spiral pattern that we use in our simulations is considered
detached from the ends of the bar. In line with the gas dynamical
model of the galaxy, \cite{bissantz}, we choose the bar pattern speed
to be more than double that of the spiral. This is corroborated by
\cite{melnik} who also suggests an upper-limit of
25kms$^{-1}$kpc$^{-1}$ on $\Omega_s$. We experiment with three
distinct ratios of $\Omega_s : \Omega_b$ - 18/55, 21/55, 25/55. 

We choose this spiral to be a 4-armed, tightly wound structure (pitch
angle = 15$^\circ$), along the lines of \cite{johnston01}. This implies
that the ILR of this 4-armed spiral ($ILR_s$) occurs outside, on top
of and inside $OLR_b$, for the choices of $\Omega_s : \Omega_b$ -
18/55, 21/55 and 25/55, respectively. Naturally, the 21/55 model 
raises interest, given that in this case, resonance overlap occurs,
ensuring global chaos. \cite{quillen} had invoked this scenario to
explain the observed stellar streams in the vicinity of the Sun.

\subsection{Some Technicalities} 
\noindent
In our scale-free disk, all lengths are expressed in units of the
co-rotation radius of the bar ($R_{CR}$). All pattern speeds are also
expressed in terms of the bar pattern speed, which is set to unity.

We choose to record orbits in an annular region between $R/R_{CR}
= 1.7$ to $R/R_{CR}=2.3$. The significance of the value of 1.7 is
that the Outer Lindblad Resonance of an $m=2$ perturbation occurs at
$R/R_{CR} = 1.7$, in a background Mestel potential.

All orbits are recorded in the frame that is rotating with the
bar. Azimuth $\phi=0$ coincides with the bar major axis. In the
bar+spiral simulations, where the figure of the spiral is not static
in the frame of the bar, we record the orbits only when, at
$R=R_{CR}$, the potential of the spiral is maximum. At the end of the
simulations, each orbit is sampled (1000 times) in time, which is
equivalent to sampling in phase, assuming ergodicity to be valid, to
provide us the output.
%About 3500 orbits (approximately the same number as the number
%of stars used to construct the local $U-V$ diagram) are each sampled
%1000 times.

\section{Results}
\noindent
In our simulations, we work with 5 models. These are:
\begin{itemize}
\item bar only model with no perturbation from the spiral: model $bar-only$.
\item bar and spiral, with $\Omega_s : \Omega_b$=18/55: model $bar+sp_{18}$.
\item bar and spiral, with $\Omega_s : \Omega_b$=21/55: model $bar+sp_{21}$.
\item bar and spiral, with $\Omega_s : \Omega_b$=25/55: model $bars+p_{25}$.
\item spiral only model with no perturbation from the bar: model $spiral-only$
\end{itemize}

The output orbits are first put on a regular polar grid; each of our
$R-\phi$ bins are chosen to be 0.025$R_{CR}$ wide in radius and
10$^\circ$ wide in azimuth. At each relevant $R-\phi$ cell, the
recorded velocities are put on a regular Cartesian $U-V$ grid. With
this stored kinematic data, velocity distributions are prepared at
each $R-\phi$ bin. When the simulated velocity distribution at a given
disk location ($f_s(R,\phi)$)is found to match the observed
distribution ($f_o$) well, we refer to such a velocity distribution as
a ``good'' distribution and the corresponding $R-\phi$ bin is called a
``good'' location.

The quality of overlap between $f_o$ and $f_s(R,\phi)$ is quantified
by a goodness-of-fit technique (discussed below). All this is
undertaken individually for each of the 5 models.

%\begin{figure}
%\hspace{-2cm}
%\includegraphics[height=19cm]{all2.ps}
%\caption{Examples of ``good'' $U-V$ distributions from 4 different
%``good'' locations, as obtained from 4 of the models we work with; top
%left: $bar_only$, top right: $bar+sp_{18}$, bottom left: $bar+sp_{25}$,
%bottom right: $bar+sp_{21}$. The observed distribution is shown in filled
%contours in the background while the simulated distributions are
%overplotted in broken lines.}
%\label{fig:1}       
%\end{figure}

\subsection{Goodness of Fit Technique}
\noindent
We test the null hypothesis: the observed $U-V$ data is drawn from
$f_s(R,\phi)$. This testing is done by estimating the $p$-value of a
test statistic $S$. $p$-value measures the probability of how unlikely
it is that a null hypothesis is true by fluke; thus, low $p$-values
indicate bad fits to the data. Though extensive literature exists to
suggest various options for $S$, we find that the definition of $S$ as
the reciprocal of the likelihood $L$, suffices in our case. Here, the
likelihood of data $D$, given the simulated distribution $f_s(R,\phi)$
is:
\begin{equation}
L(D|f_s) = \prod\nu_{R,\phi}(U_i,V_j), 
\end{equation}
where $\nu_{R,\phi}(U_i,V_j)$ is the value of $f_s(R, \phi)$, in the
$(i,j)^{th}$ $U-V$ bin and the product is performed over all the $U$
and $V$ velocity bins. We decide to choose the pre-set significance
level of 5$\%$, for the acceptance of the null hypothesis.

\subsection{Identification of ``Good'' Locations}
\noindent
For a given model, the quality of overlap between $f_o$ and
$f_s(R,\phi)$ is given by the $p$-value of the test statistic in that
$R-\phi$ bin. This is referred to as $p-value(R, \phi)$. Distributions
of $p-value(R, \phi)$ are shown in Figure~1, for models $bar-only$ and
$spiral-only$. It is to be noted that the function $p-value(R, \phi)$
for the resonance overlap ($bar+sp_{21}$) and $spiral-only$ models is
distinct from that for the other three models. This is attributed to
the chaos inducing efficacy of the spiral perturbation.

\begin{figure}
\hspace{-1cm}
\includegraphics[height=6cm]{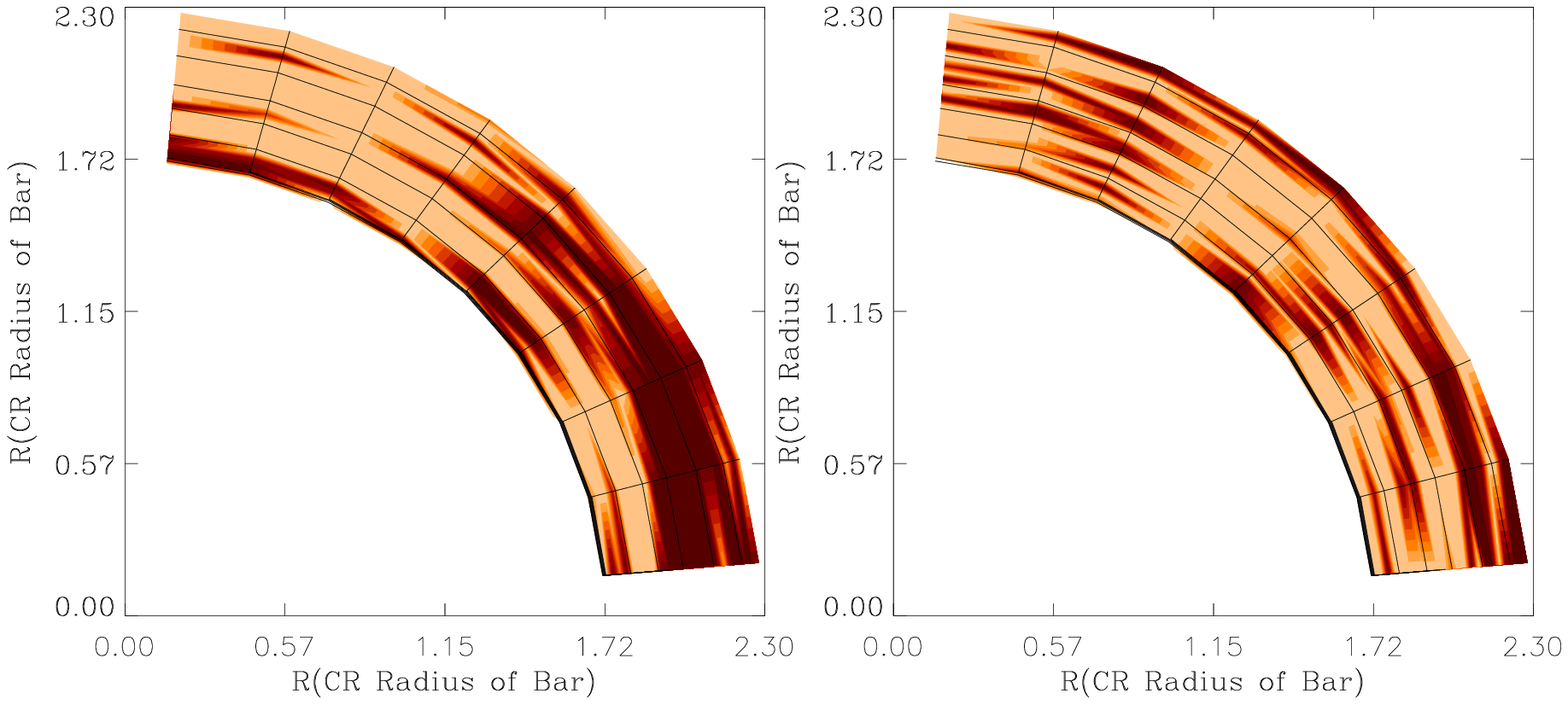}
\caption{The distribution $p-value(R, \phi)$ over the part of the
annulus that we record our orbits in, for models $bar-only$ (left)
and $spiral-only$ (right). The $R-\phi$ cells in the darker colours
indicate higher $p$-values than those in lighter shading.} 
\label{fig:1}       
\end{figure}

\begin{figure}
\hspace{-1cm}
\includegraphics[height=7cm]{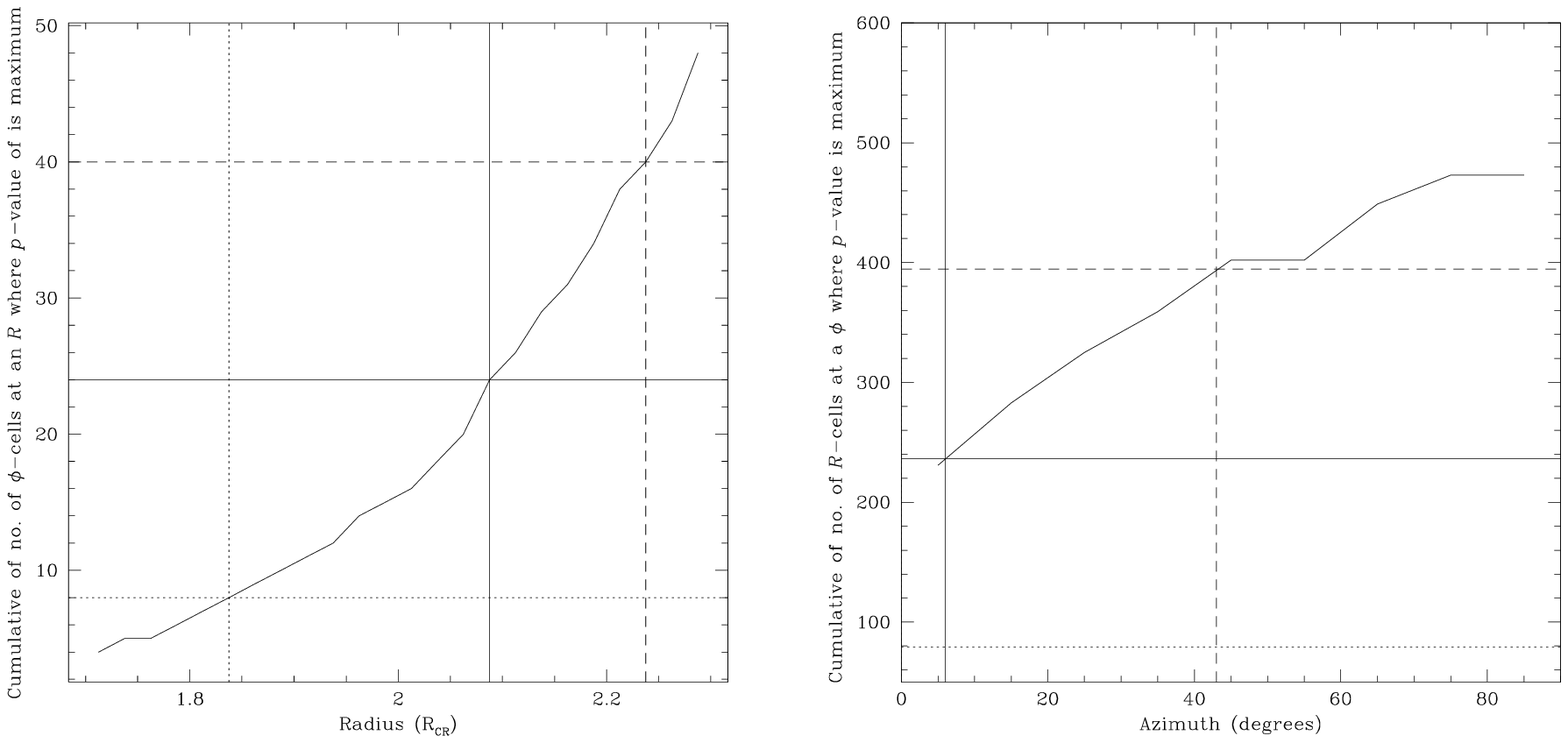}
\caption{The distribution of the maximal $p$-value locations is
marginalised, over azimuth and radius, to yield best-fit values of
observer radial and azimuthal location (respectively), in order for
the observer's local velocity distribution to best match the observed
distribution.}
\label{fig:2}       
\end{figure}

From the distribution $p-value(R, \phi)$, ``good'' locations are
identified as those $R-\phi$ addresses, at which the $p$-value attains
the maximum, i.e. 100$\%$. This distribution of the ``good'' locations
can be marginalised over $\phi$ to allow us to quantify the
$\pm$1-$\sigma$ range that defines the best radial
locations. Similarly, marginalising over $R$ gives the constraints on
the best azimuthal location for the observer, in a frame where
$\phi$=0 implies bar major axis.  Of course, these are then
constraints on our location, i.e. the Solar position, in the frame of
the bar. Such marginalised distributions (with $R$ and $\phi$) are
shown in Figure~2, for model $bar+sp_{25}$. This tells us that
\begin{itemize}
\item the best radial location for the observer is $R/R_{\rm CR}$=
2.0875$^{0.15}_{0.25}$. Equating the Solar radius (of 8 kpc) with the
medial value, we scale the bar pattern speed to
57.4$^{4.1}_{6.9}$\kmskpc (using circular velocity at Sun=220 km s$^{-1}$).
\item the best fit values for the bar angle is [0$^\circ$,
43$^\circ$], with a medial value of 6$^\circ$.
\end{itemize}

We find that all the models that include the influence of the bar and
the spiral pattern are able to produce the observed stellar streams in
the vicinity of the Sun, at different disk locations. Though the
$bar-only$ model is also capable of the same, it has to be discarded
in light of the dynamical reasoning that no model of the Solar
neighbourhood is complete without taking the effect of the spiral into
account (\cite{chakrabarty_aa}). The $spiral-only$ model is rejected
since it fails to warm up the local patch in the disk enough, to
guarantee velocity dispersions of the order of those observed in the
Solar neighbourhood.

\section{Origin of the Splitting of the Larger Mode}
\noindent
On the basis of the results above, we conclude that the splitting of
the larger mode in the local $U-V$ distribution is not due to
resonance overlap since bar+spiral models that do not impose resonance
overlap are also successful in producing the observed phase space
structure. On the other hand, this splitting is suggested to be due to
the interaction of chaotic orbits and orbits that belong to families
corresponding to the minor resonances due to the spiral and the bar
(such as the outer $1:1$ resonance of the bar).

\begin{figure}
\centering
\includegraphics[height=8cm]{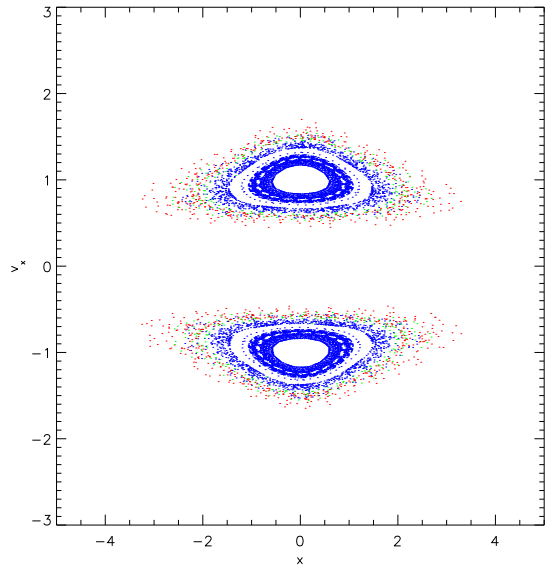}
\caption{Surface of section of orbits at effective energy value of
-1, from the $bar+sp_{18}$ model. The strongly chaotic orbits are
in medium grey shading, while the regular orbits are in the darkest
shade. The intermediate shading corresponds to the weakly chaotic
orbits.}
\label{fig:3}       
\end{figure}

As for the overall bimodal character of the local phase space, a
rudimentary orbital analysis confirms that in line with the Kalnajs
Mechanism, the overall bimodality is indeed due to scattering off
$OLR_b$; the Hercules stream appears to be built of quasi-periodic
orbits belonging to the anti-aligned family while the anti-aligned
family was not spotted in the other larger mode. (The other mode
manifests orbits from the aligned family).

\section{The Role of Chaos}
\noindent
\cite{fux_aa} has suggested bar-induced chaos as the cause for the local
streams. However, in a chaos quantification exercise that is currently
underway (Chakrabarty $\&$ Sideris, {\it submitted to A$\&$A}), the
orbits in the $bar-only$ model are found to be regular; it is noted
that it is the presence of the spiral that triggers the onset of
chaos. Figure~3 indicates the relative fraction of chaotic orbits,
over regular and weakly chaotic orbits, in the $bar+sp_{18}$
models. Similar surfaces of section for the $bar-only$ models, at even
higher energies indicate no irregularity.

It is of course possible that a stronger bar would induce greater
irregularity than in our work, but the point is that the local phase
space structure can be emulated with the bar strength used herein.

\section{Summary $\&$ Future Work}
\noindent
In this contribution, we have attempted to understand the state of the
local phase space via dynamical modelling that includes the influence
of the bar and the spiral pattern in the Galaxy. The clumpy structure
of the 2-D local velocity distribution was attributed to the major and
minor resonances of these perturbations, as well as the emergence of
chaoticity triggered primarily by the spiral potential. Only the
models that take the gravitational potentials of both bar and
spiral are found to offer viable representations of the Solar
neighbourhood dynamics; overlap of the results from the successful
models indicate a fast bar, with a pattern speed of
57.4$^{+3.0}_{-2.5}$\kmskpc and a bar angle that lies in the range
[0$^\circ$, 30$^\circ$].

The currently used $p$-value formalism is not satisfactory on account
of the lack of automativeness and inability to rank or grade the
models. In fact, we are hoping to improve upon this technique, with a
Bayesian measure of the quality of a simulation. This is to aid the
expansion of the relevant parameter space, particularly, in the
inclusion of the effects of the halo and progression to a full six
dimensional phase space.

%\bibliographystyle{}
%\bibliography{}

\printindex
\end{document}